\documentclass[aps,prl,amsmath,twocolumn,superscriptaddress,floatfix,footinbib,showpacs,longbibliography]{revtex4-1}

\usepackage{graphicx}
\usepackage{epstopdf}

\usepackage[T1]{fontenc}
\usepackage[applemac]{inputenc}
\usepackage{lmodern}
\usepackage[english]{babel}

\usepackage{ae}
\usepackage{units}

\usepackage{amsmath,amssymb,natbib,bm}
\usepackage{psfrag}
\usepackage{subfigure}
\usepackage{amsthm}

\usepackage{fixmath}
\usepackage{booktabs}

\usepackage{slashed}

\usepackage[americaninductors]{circuitikz}
\usepackage{tikz}
\usetikzlibrary{arrows}

\usepackage{color}
\usepackage{url}

\usepackage[colorlinks]{hyperref}
\hypersetup{%
        plainpages=true,
        breaklinks=true,
        hypertexnames=false,
        pageanchor=true,
        colorlinks=true,
        linkcolor={blue},
        citecolor={magenta},
        urlcolor={blue},
        anchorcolor={black}
      }

\newcommand{\figref}[1]{\mbox{Fig.~\ref{#1}}}
\newcommand{\tabref}[1]{\mbox{Table~\ref{#1}}}

\renewcommand{\eqref}[1]{\mbox{Eq.~(\ref{#1})}}

\newcommand{\rd}{\ensuremath{\mathrm{d}}}
\newcommand{\id}{\ensuremath{\,\rd}}

\newcommand{\ket}[1]{|#1\rangle}

\newcommand{\ketbra}[2]{\left| #1 \rangle \langle #2 \right|}

\newcommand{\expec}[1]{\left\langle #1 \right\rangle}

\newcommand{\comm}[2]{\left[ #1, #2 \right]}
\newcommand{\lind}[1]{\mathcal{D}\left[#1\right]}

\newcommand{\abs}[1]{\left|#1\right|}

\newcommand{\nn}{\nonumber}

\newcommand{\be}{\begin{equation}}
\newcommand{\ee}{\end{equation}}
\newcommand{\bea}{\begin{eqnarray}}
\newcommand{\eea}{\end{eqnarray}}

\begin{document}

\title{Reflective amplification without population inversion from a strongly driven superconducting qubit}

\author{P.~Y.~Wen}
\affiliation{Department of Physics, National Tsing Hua University, Taiwan}

\author{A.~F.~Kockum}
\affiliation{Center for Emergent Matter Science, RIKEN, Saitama 351-0198, Japan}

\author{H.~Ian}
\affiliation{Institute of Applied Physics and Materials Engineering, University of Macau, Macau}
\affiliation{UMacau Zhuhai Research Institute, Zhuhai, Guangdong, China}

\author{J.~C.~Chen}
\affiliation{Department of Physics, National Tsing Hua University, Taiwan}

\author{F.~Nori}
\affiliation{Center for Emergent Matter Science, RIKEN, Saitama 351-0198, Japan}
\affiliation{Physics Department, The University of Michigan, Ann Arbor, Michigan 48109-1040, USA}

\author{I.-C.~Hoi}
\email[e-mail:]{ichoi@phys.nthu.edu.tw}
\affiliation{Department of Physics, National Tsing Hua University, Taiwan}

\date{\today}

\begin{abstract}

Amplification of optical or microwave fields is often achieved by strongly driving a medium to induce population inversion such that a weak probe can be amplified through stimulated emission. Here we strongly couple a superconducting qubit, an artificial atom, to the field in a semi-infinite waveguide. When driving the qubit strongly on resonance such that a Mollow triplet appears, we observe a 7\% amplitude gain for a weak probe at frequencies in-between the triplet. This amplification is not due to population inversion, neither in the bare qubit basis nor in the dressed-state basis, but instead results from a four-photon process that converts energy from the strong drive to the weak probe. We find excellent agreement between the experimental results and numerical simulations without any free fitting parameters. The device demonstrated here may have applications in quantum information processing and quantum-limited measurements.

\end{abstract}


\maketitle

Superconducting qubit circuits are playing an important role in the development of solid-state quantum computation~\cite{You2005, Devoret2013, Wendin2016}, where they already have been used in implementations of quantum logic gates and algorithms~\cite{Lucero2012, Fedorov2012}. However, in the past decade, superconducting qubit circuits have also become a prominent platform for quantum-optics research. This development originated from the achievement of strong coupling in the circuit version of cavity quantum electrodynamics, where a superconducting qubit acts as a substitute for the atom and a strip-line waveguide replaces the optical cavity~\cite{Wallraff2004}. In this context, a broad range of phenomena from atomic physics and quantum optics~\cite{You2011, Gu2017}, e.g., lasing~\cite{Astafiev2007, Ashhab2009, You2007}, have been demonstrated in solid-state systems. Some of these phenomena, e.g., electromagnetically induced transparency~\cite{Abdumalikov2010, Joo2010, Ian2010, Peng2014, Sun2014, Liu2016}, can be demonstrated with greater clarity and sophistication than in corresponding experiments with natural atoms.

The additional capabilities in quantum optics with superconducting circuits stems partly from the tunable and designable nature of the superconducting qubits as two-level (or multi-level) systems, and partly from the ease with which strong coupling can be achieved between these artificial atoms and quantum fields. The latter property has permitted the demonstration of ultrastrong coupling~\cite{Bourassa2009, Niemczyk2010, Yoshihara2017, Chen2017}, going beyond the physics of the rotating-wave approximation and the Jaynes--Cummings model, and the dressed Zeno effect~\cite{Li2014} with the coupling taking place in a resonator. With the coupling being between a superconducting qubit and an open waveguide~\cite{Astafiev2010, Hoi2013b, VanLoo2013, Fang2015, Roy2017, Forn-Diaz2017} instead, it has made possible the demonstration of, e.g., the Mollow triplet~\cite{Mollow1969, Astafiev2010, Abdumalikov2011} and single-photon routing~\cite{Hoi2011}. If one truncates the open waveguide, forming a semi-infinite space with the qubit placed close to the endpoint (equivalent to placing an atom in front of a mirror~\cite{Hoi2015}), the qubit becomes strongly coupled to a single input-output channel. Such a setup has been used to demonstrate a giant cross-Kerr phase shift with a probe and a signal field interacting with different transitions in a three-level artificial atom~\cite{Hoi2013a}.

In this Letter, we leverage the excellent characteristics of a superconducting qubit at the end of a transmission line to demonstrate amplification in a two-tone experiment. We find that, when a strong resonant drive field splits the qubit transition into a Mollow triplet, a weak probe field is amplified if it is tuned to the frequencies between the resonances in that triplet. This is in accordance with another theoretical prediction by Mollow~\cite{Mollow1972}. Previous experiments with many natural atoms~\cite{Wu1977} and a single quantum dot~\cite{Xu2007} have reported similar amplification, but only at levels of 0.4\% and 0.005\%, respectively. In our experiment, we measure amplitude gain reaching up to 7\%. We note that the amplification mechanism does not rely on population inversion, as in another experiment with a superconducting three-level artificial atom in an open waveguide~\cite{Astafiev2010a}, nor even population inversion between dressed states, as in another such experiment~\cite{Koshino2013} and some implementations of lasing without inversion~\cite{Mompart2000}. Instead, the amplification can be explained in terms of higher-order processes with stimulated emission and transitions between dressed states~\cite{Friedmann1987}.

\begin{figure}
\includegraphics[width=\linewidth]{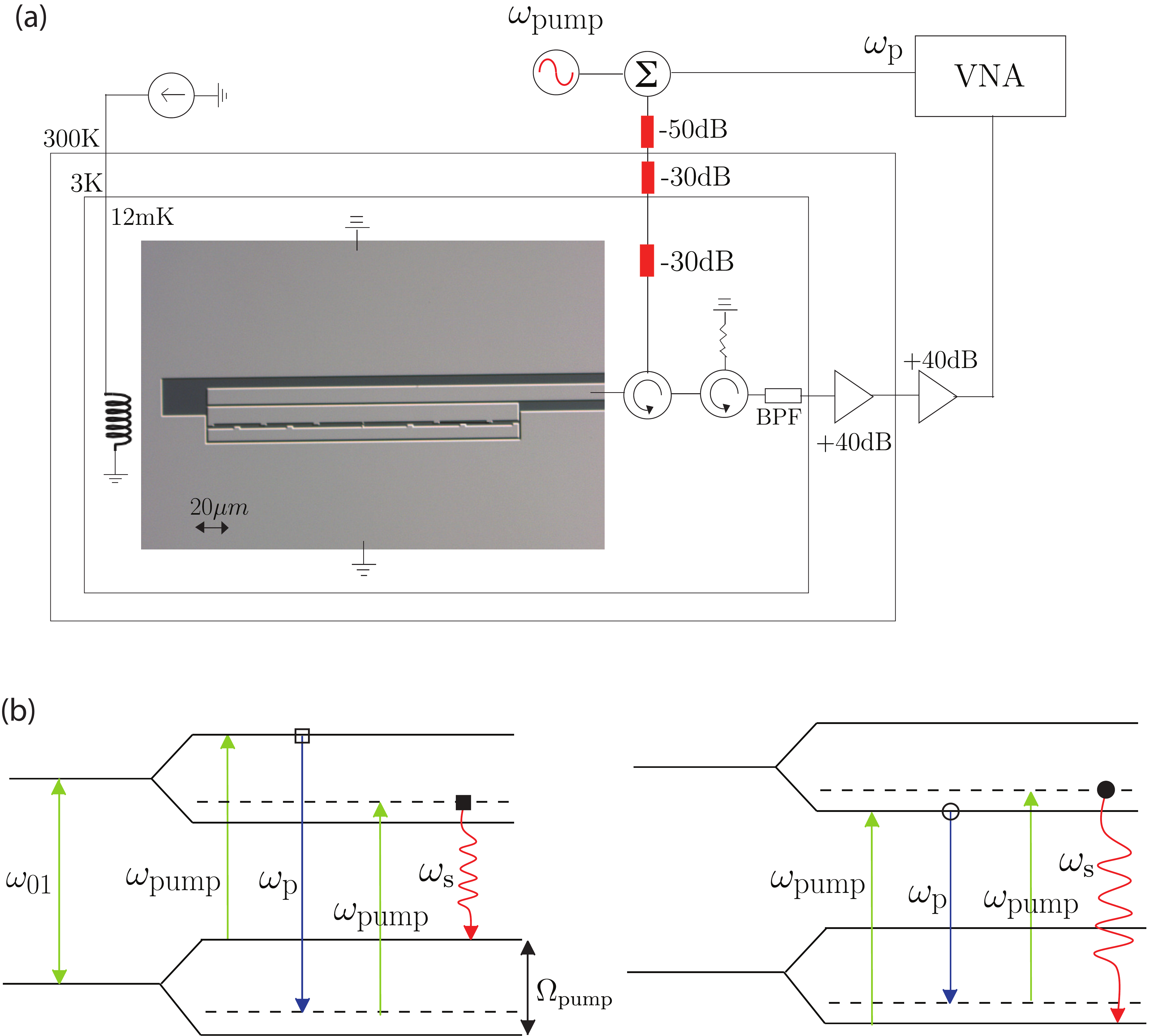}
\caption{Experimental setup and amplification mechanism. (a) A simplified schematic of the device and the setup for the experiment. The transmon qubit, our artificial atom, is formed by two superconducting islands (center of the image) coupled through two Josephson junctions and a large capacitance. The qubit sits at the end of the transmission line formed by the center conductor and ground planes pictured here. The microwave pump and probe tones are generated at room temperature, combined at $\Sigma$, and fed through attenuators to the qubit in a cryostat cooled to $\unit[12]{mK}$. The output signal is measured in a vector network analyzer (VNA). (b) Two energy-level diagrams showing how the weak probe is amplified in our setup. The strong pump with amplitude $\Omega_{\rm pump}$ dresses the energy levels of the bare two-level qubit. The left part shows one of the two irreversible four-photon processes that leads to the stimulated emission of a photon at the probe frequency $\omega_{\rm p}$ (blue arrow) when $\omega_{\rm pump} + \Omega_{\rm pump} > \omega_{\rm p} > \omega_{\rm pump} + \sqrt{2\Gamma_1\gamma^3} / \Omega_{\rm pump}$, and the right part shows the same for the case $\omega_{\rm pump} - \sqrt{2\Gamma_1\gamma^3} / \Omega_{\rm pump} > \omega_{\rm p} > \omega_{\rm pump} - \Omega_{\rm pump}$. In these processes, two pump photons (green arrows) are absorbed and an additional photon (red wavy arrow) is scattered at frequency $\omega_{\rm s} = 2\omega_{\rm pump} - \omega_{\rm p}$. An important part of the processes is that the two virtual states shown with dashed lines are connected by the strong pump. This increases the probability of the process occurring, since these virtual states are within the power-broadening width of the pump. The empty and filled circles and squares are used to mark the corresponding transitions in \figref{fig:PumpPower}(b). 
\label{fig:Device}}
\end{figure}

The device used in our experiment is shown in \figref{fig:Device}(a). A transmon qubit~\cite{Koch2007} is embedded at the end of a one-dimensional transmission line with characteristic impedance $Z_0 \simeq \unit[50]{\Omega}$. We denote the ground state, the first excited state, and the second excited state of the transmon by $\ket{0}$, $\ket{1}$, and $\ket{2}$, respectively. The $\ket{0} \leftrightarrow \ket{1}$ transition energy of the transmon is $\hbar \omega_{10}(\Phi) \approx \sqrt{8 E_J(\Phi) E_C} - E_C$; it is determined by the charging energy $E_C = e^2 / 2C_\Sigma$, where $e$ is the elementary charge and $C_\Sigma$ is the total capacitance of the transmon, and the Josephson energy $E_J(\Phi) = E_J \abs{\cos(\pi \Phi / \Phi_0)}$. The Josephson energy can be tuned from its maximum value $E_J$ by the external flux $\Phi$ of a magnetic coil; $\Phi_0 = h / 2e$ is the magnetic flux quantum. Due to the position of the transmon at the end of the transmission line, the field emitted from the transmon can only propagate in one direction. The diagram in \figref{fig:Device}(a) also illustrates the rest of the experimental setup. The pump field at frequency $\omega_{\rm pump}$ and the probe field of frequency $\omega_{\rm p}$ are fed into the transmission line via a combiner and several attenuators. The output signal is amplified and measured in a vector network analyzer (VNA) to determine the amplitude reflection coefficient $r$ of the probe field.

The origin of the amplification is illustrated in \figref{fig:Device}(b), following Ref.~\cite{Friedmann1987}. The two-level structure of the qubit becomes dressed by the strong resonant ($\omega_{\rm pump} = \omega_{10}$) pump with Rabi frequency $\Omega_{\rm pump}$, forming the dressed states which have transitions at the Mollow-triplet frequencies $\omega_{10}$ and $\omega_{10} \pm \Omega_{\rm pump}$. Since the drive is resonant, all states are equally populated; no population inversion occurs. When $\omega_{\rm p} = \omega_{10} \pm \Omega_{\rm pump}$, the equal population of all states leads to absorption and stimulated emission being equally likely, which means that the probe experiences neither gain nor attenuation. If $\abs{\omega_{\rm p} - \omega_{10}} > \Omega_{\rm pump}$, a three-photon process that leads to absorption dominates. However, when $\sqrt{2\Gamma_1\gamma^3} / \Omega_{\rm pump} < \abs{\omega_{\rm p} - \omega_{10}} < \Omega_{\rm pump}$~\cite{Mollow1972}, four-photon processes like the ones shown in \figref{fig:Device}(b) dominate and lead to amplification. Here, $\Gamma_1$ is the relaxation rate and $\gamma$ is the decoherence rate for the qubit.

\begin{figure}
\includegraphics[width=\linewidth]{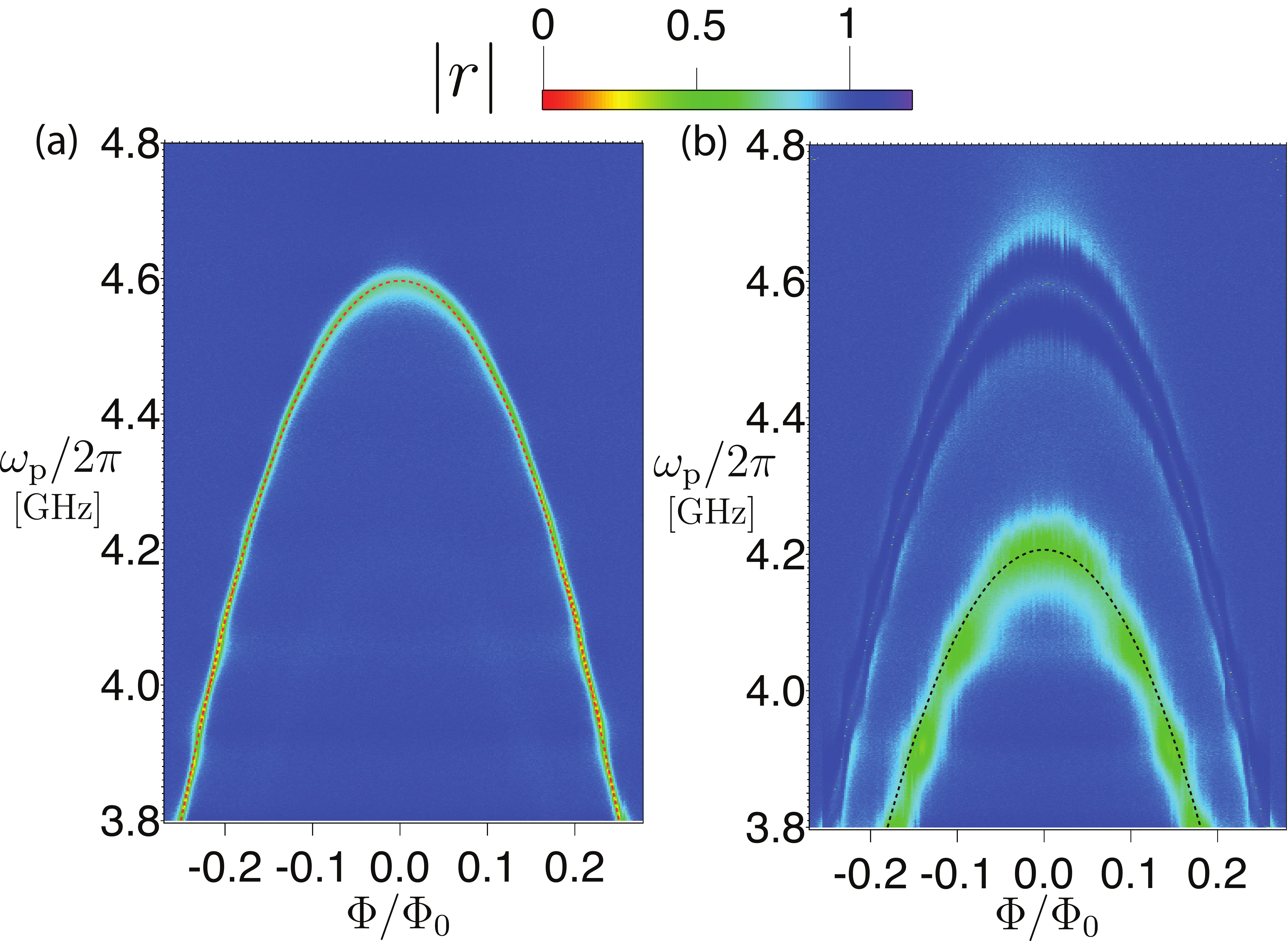}
\caption{Spectroscopy of the transmon qubit. (a) Single-tone spectroscopy. Amplitude reflection coefficient $\abs{r}$ as a function of probe frequency $\omega_{\rm p}$ and flux $\Phi$. The red dashed curve is a fitted theory curve for $\omega_{10}$. (b) Two-tone spectroscopy. We once again show $\abs{r}$ for the weak probe as a function of $\omega_{\rm p}$ and $\Phi$, but in this experiment a second, strong microwave drive is applied at $\omega_{10}$. The broad green region that appears in the response corresponds to the $\ket{1} \leftrightarrow \ket{2}$ transition at frequency $\omega_{21}$. The black dashed curve is a fitted theory curve for $\omega_{21}$. We also see the features of the Mollow triplet around $\omega_{10}$ and note that the features of this triplet can be tuned by the flux.
\label{fig:SpectroscopyParabolas}}
\end{figure}

Before performing the amplification experiments, we first characterize our device spectroscopically. In \figref{fig:SpectroscopyParabolas}(a), we show the amplitude reflection coefficient $\abs{r}$ of a weak probe (amplitude $\Omega_{\rm p} \ll \gamma$) as a function of the external flux $\Phi$. We clearly see the $\Phi$ dependence of the $\omega_{10}$ transition frequency. The transmon also has higher levels. To see the next transition, between $\ket{1}$ and $\ket{2}$ at frequency $\omega_{21}$, we use two-tone spectroscopy. We saturate the $\ket{0} \leftrightarrow \ket{1}$ transition by applying a pump field at $\omega_{10}$ with pump power $P_{\rm pump} = \unit[-119]{dBm}$, and measure the reflection of a weak probe at $\omega_{\rm p}$. As can be seen in \figref{fig:SpectroscopyParabolas}(b), we observe photon scattering from the $\ket{1} \leftrightarrow \ket{2}$ transition, which appears as a dip in the reflection at $\omega_{\rm p} = \omega_{21}$. We also observe that the strong resonant pump dresses the $\ket{0} \leftrightarrow \ket{1}$ transition, giving rise to three resonances around $\omega_{10}$. This is the well-known Mollow triplet, which we study further in the amplification experiments below.

\begin{table*}
\centering
\begin{tabular}{| c |c | c | c | c | c | c | c | }
\hline
 $E_{J,0} / h$ [GHz] & $E_C / h$ [GHz] & $E_{J,0} / E_C$ & $\omega_{10} / 2\pi$ [GHz] & $\omega_{21} / 2\pi$ [GHz] & $\Gamma_1 / 2\pi$ [MHz] & $ \Gamma_\phi / 2\pi$ [MHz] & $\gamma / 2\pi$ [MHz] \\
\hline
 $7.97$ & $0.39$ & $20.4$ & $4.59$ & $4.2$ & $45$ & $2.7$ & $25.2$ \\
\hline
\end{tabular}
\caption{Extracted parameters of the device.
\label{tab:Params}}
\end{table*} 

From Figs.~\ref{fig:SpectroscopyParabolas}(a,b), we extract $E_J = \unit[7.97]{GHz}$, $E_C = \unit[390]{MHz}$, $\omega_{10}(\Phi)$, and $\omega_{21}(\Phi)$. We then perform further single-tone scattering experiments at $\Phi / \Phi_0 = 0$ as in Ref.~\cite{Hoi2013a}. From the magnitude and phase of the reflection coefficient for a weak probe as a function of $\omega_{\rm p}$, we extract $\Gamma_1$, the pure dephasing $\Gamma_\phi$, and $\gamma$. Measuring $r$ as a function of the probe power $P$, we extract the coupling constant $k$ relating the input power to the Rabi frequency according to $\Omega_p / 2\pi = k \sqrt{P}$. All the extracted parameters are summarized in \tabref{tab:Params}. We note that $\Gamma_1$ is dominated by the coupling to the transmission line and greatly exceeds $\Gamma_\phi$, placing our system in the strong-coupling regime.

\begin{figure}
\includegraphics[width=\linewidth]{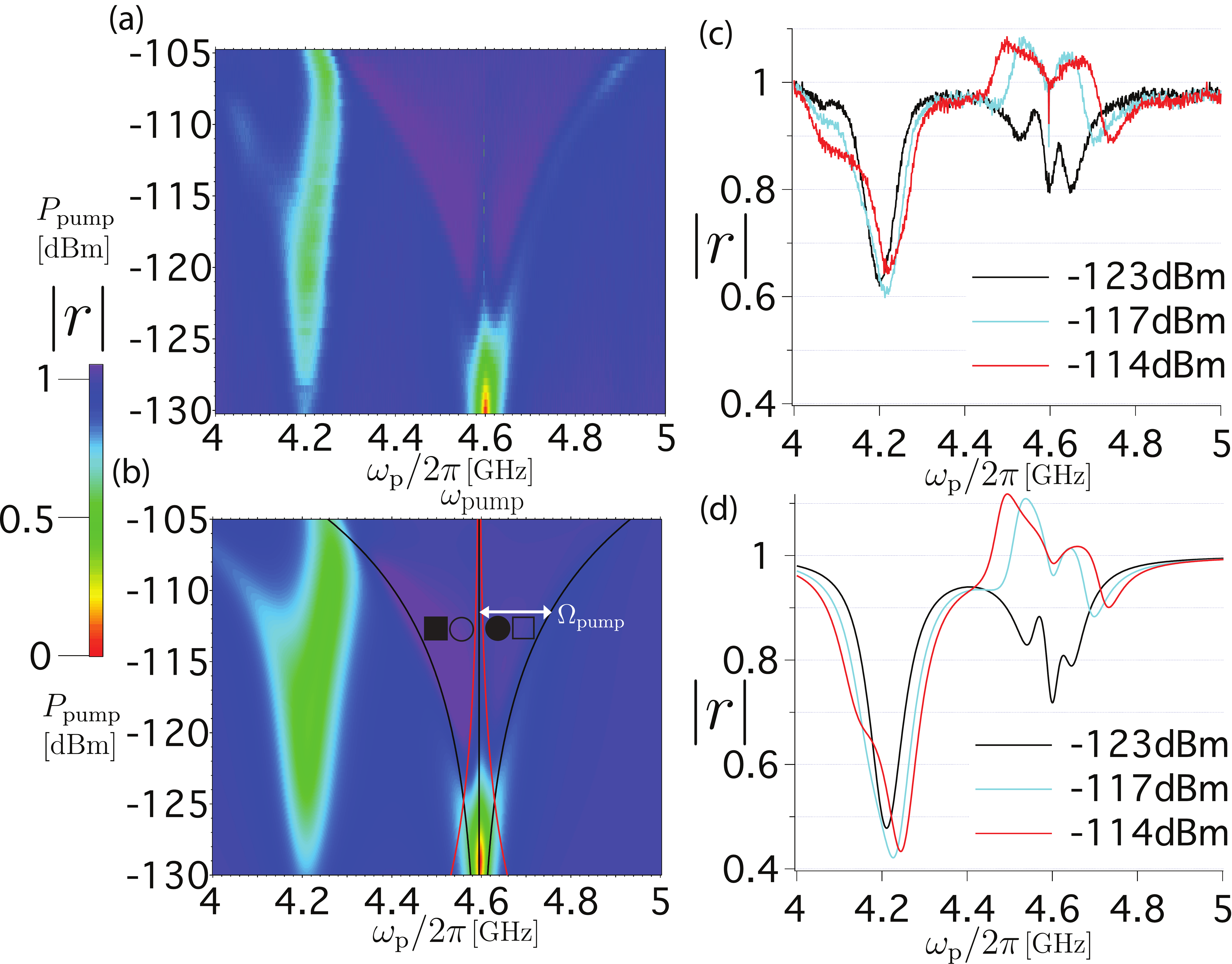}
\caption{Gain and attenuation around the Mollow triplet. (a) Reflection coefficient $\abs{r}$ as a function of probe frequency and pump power at $\Phi = 0$. The pump frequency $\omega_{\rm pump}$ is fixed at $\unit[4.59]{GHz}$, which is the transition frequency $\omega_{10}$. (b) A numerical simulation of the experiment using the parameters in \tabref{tab:Params} and $N=5$ energy levels for the transmon. The solid black curves indicate the position of the Mollow triplet at $\omega_{\rm pump}$ and $\omega_{\rm pump} \pm \Omega_{\rm pump}$. The red curves correspond to the expected inner amplification boundaries $\omega_{\rm pump} \pm \sqrt{2\Gamma_1\gamma^3} / \Omega_{\rm pump}$. The empty and filled circles and squares correspond to the frequencies for stimulated emission and scattering for the four-photon processes sketched in \figref{fig:Device}(b). (c) Three line cuts at different pump powers: $\unit[-123]{dBm}$, $\unit[-117]{dBm}$, and $\unit[-114]{dBm}$. We observe three dips around $\omega_{10}$, close to the Mollow triplet resonances. We also observe two amplification peaks in-between these dips. The gain, which reaches about 7\% for $P_{\rm pump} = \unit[-114]{dBm}$, is due to stimulated emission in higher-order processes as explained in \figref{fig:Device}(b). We note again that the amplification is not due to population inversion. (d) The corresponding line cuts from the numerical simulation.
\label{fig:PumpPower}}
\end{figure}

We now investigate the Mollow-triplet structure further by fixing the flux at $\Phi = 0$ and the pump frequency at $\omega_{\rm pump} = \omega_{10}$. We record the reflection coefficient $\abs{r}$ of a weak probe as a function of both $\omega_{\rm p}$ and pump power, increasing $P_{\rm pump}$ from $\unit[-130]{dBm}$ to $\unit[-105]{dBm}$. The result is shown in \figref{fig:PumpPower}(a). To the left in the figure, around $\omega_{\rm p} = \unit[4.2]{GHz}$, we see scattering from the $\ket{1} \leftrightarrow \ket{2}$ transition, which becomes possible because the resonant pump populates the first excited state of the qubit. As the pump power increases, this feature splits into two dips in the reflection, which corresponds to an Autler-Townes doublet~\cite{Autler1955, Sun2014, Liu2016}. Increasing the pump power, we also see the resonance around $\omega_{10}$ separate into a Mollow-triplet-like structure. The central transition becomes weaker at higher powers (the delta peak/dip in the data exactly at $\omega_{10}$ is an artifact of the pump, unrelated to the qubit response) and areas with greater-than-unity reflection, $\abs{r} > 1$ (purple color), appear in-between the three dips. This is further illustrated in \figref{fig:PumpPower}(c), which shows three line cuts at different pump powers. The two reflection peaks correspond to gain arising from conversion between pump and probe photons, mediated by the superconducting artificial atom, as explained in \figref{fig:Device}(b). We emphasize that this gain is not a result of population inversion, not even among the dressed states. The maximum gain we observe is about 7\%, at $P_{\rm pump} = \unit[-114]{dBm}$. We note that the two gain peaks are asymmetric, unlike what Mollow predicted for a two-level atom. We attribute this asymmetry to influence from the second excited state of the transmon qubit.

\begin{figure*}
\includegraphics[width=\linewidth]{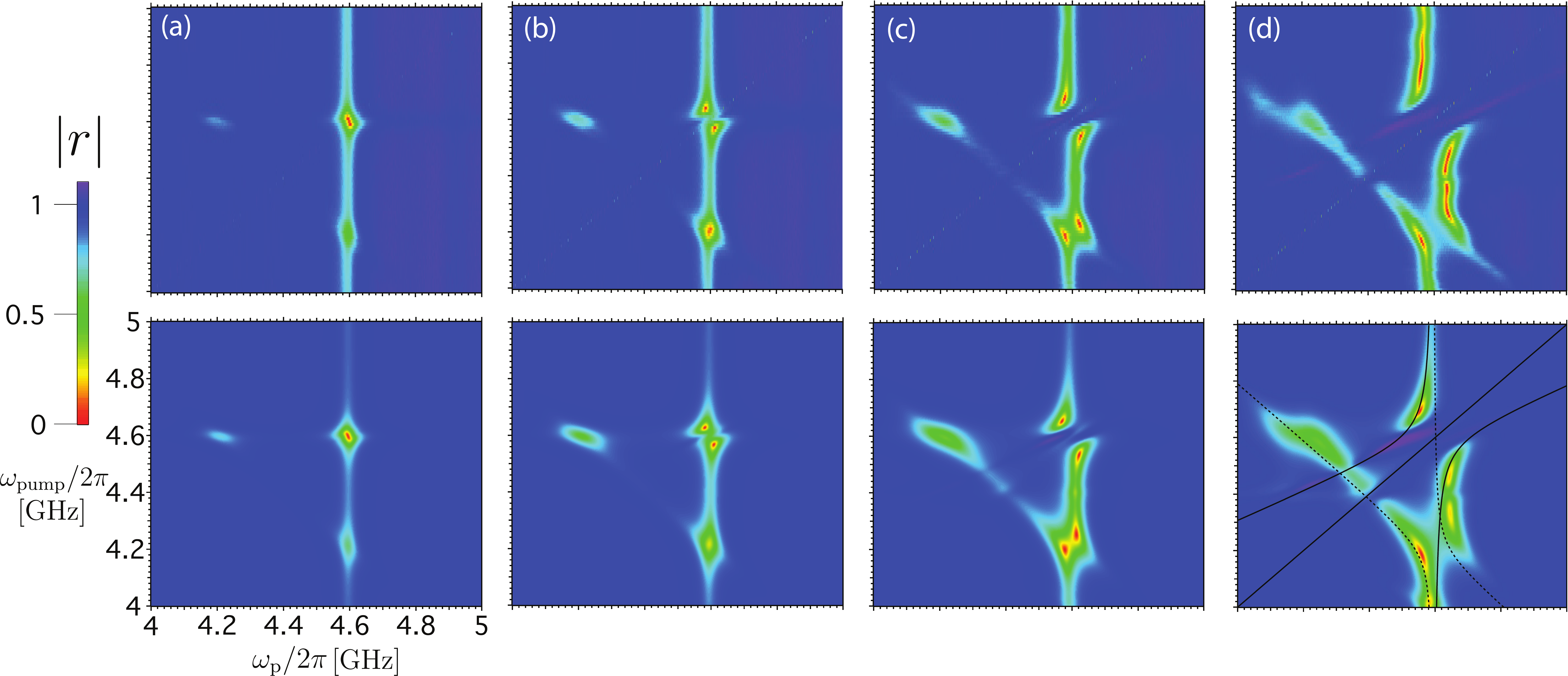}
\caption{Reflection coefficient $\abs{r}$ of a weak probe as a function of $\omega_{\rm p}$ ($x$ axis) and $\omega_{\rm pump}$ ($y$ axis) at $\Phi = 0$ for four different pump powers: (a) $\unit[-130]{dBm}$, (b) $\unit[-125]{dBm}$, (c) $\unit[-120]{dBm}$, and (d) $\unit[-115]{dBm}$. The top row is experimental data and the bottom row is numerical simulations, performed in the same way as in \figref{fig:PumpPower} using the parameters in \tabref{tab:Params} (no free fitting parameters). As we increase the pump power, the features at $\omega_{10} = \unit[4.59]{GHz}$ and $\omega_{21} = \unit[4.2]{GHz}$ split into a more complicated response, which can be explained in terms of the Mollow triplet due to pumping of the $\ket{0} \leftrightarrow \ket{1}$ transition [solid black lines in (d)] and the Autler-Townes doublet due to pumping of the $\ket{1} \leftrightarrow \ket{2}$ transition [dashed black lines in (d)].
\label{fig:TwoToneFrequencySweeps}}
\end{figure*}

In Figs.~\ref{fig:PumpPower}(b,d), we show numerical simulations corresponding to the experimental results in Figs.~\ref{fig:PumpPower}(a,c), respectively. These theoretical plots are done using the parameters in \tabref{tab:Params}. Even though no free fitting parameters are used, we see excellent agreement between theory and experiment. To model the two-tone spectroscopy results, we solve the master equation
\bea
&&\dot{\rho} = - \frac{i}{\hbar} \comm{\sum_{m=0}^{N-1} \hbar \Delta_m \ketbra{m}{m} - i \frac{\Omega_{\rm pump}}{2} \left( \Sigma_+ - \Sigma_- \right)}{\rho} \nn\\
&&+ \sum_{m=1}^{N-1} m \Gamma_1 \lind{\ketbra{m-1}{m}} \rho + \Gamma_\phi \lind{\sum_{m=1}^{N-1} m \ketbra{m}{m}} \rho, \quad\:\:
\eea
where $\rho$ is the density matrix for the qubit with energy levels $\Delta_m$ in the rotating frame of the pump, $\Sigma_- = \sum_{m=1}^{N-1} \sqrt{m} \ketbra{m-1}{m}$, $\Sigma_+ = \Sigma_-^\dag$, $\lind{X}\rho = X \rho X^\dag - \frac{1}{2} \left\{X^\dag X, \rho\right\}$, and $N$ is the number of qubit levels. We then use quantum linear-response theory~\cite{Kubo1957} to calculate the susceptibility
\be
\chi (\omega_{\rm p}) = i \int_0^\infty \id t \: e^{i \omega_{\rm p} t} \expec{\comm{\Sigma_-(t)}{\Sigma_p(0)}},
\ee
where $\Sigma_p = -i\left(\Sigma_+ - \Sigma_- \right)$, average over pump phases, and extract the reflection coefficient
\be
r = 1 + \Gamma_1 \chi(\omega_{\rm p}),
\ee
similar to Refs.~\cite{Kockum2013, Gustafsson2014}. The numerical simulations use methods from Ref.~\cite{Rau2004} and are implemented in QuTiP~\cite{Johansson2012, Johansson2013}.

In \figref{fig:TwoToneFrequencySweeps}, we investigate the effect of pump detuning. We sweep the frequencies of both a weak probe and a strong pump, changing the pump power in steps from $\unit[-130]{dBm}$ in \figref{fig:TwoToneFrequencySweeps}(a) to $\unit[-115]{dBm}$ in \figref{fig:TwoToneFrequencySweeps}(d). The agreement between the experimental data in the top row of \figref{fig:TwoToneFrequencySweeps} and the numerical simulations in the bottom row, performed without any free fitting parameters, is excellent. In the left parts of \figref{fig:TwoToneFrequencySweeps}, we see expected resonances when $\omega_p = \omega_{10} = \unit[4.59]{GHz}$ and when $\omega_{\rm pump} = \omega_{10}, \omega_p = \omega_{21} = \unit[4.2]{GHz}$. As pump power increases in the right parts of the figure, we observe features of the Mollow triplet when pumping at $\unit[4.59]{GHz}$ [solid black lines in \figref{fig:TwoToneFrequencySweeps}(d)] and of an Autler-Townes doublet when pumping at $\unit[4.2]{GHz}$ [dashed black lines in \figref{fig:TwoToneFrequencySweeps}(d)]. At high pump power, we also observe regions of gain (purple color). When the pump is resonant with $\omega_{10}$, the gain mechanism is the one discussed in Figs.~\ref{fig:Device} and \ref{fig:PumpPower}. When the pump is off resonance, we see gain close to one of the two sidebands of the Mollow triplet. Such gain is due to population inversion among the dressed states~\cite{Mompart2000}, making it easier to achieve than the inversionless amplification we demonstrated above, as evidenced by experiments with natural atoms~\cite{Wu1977,Khitrova1988} and with a superconducting qubit in an open transmission line~\cite{Koshino2013}. 

In summary, we have demonstrated amplification of a weak probe by about $7\%$ using a single resonantly pumped superconducting artificial atom placed at the end of a transmission line. The amplification does not rely on population inversion, not even in the dressed-state basis. The device demonstrated here may have applications in quantum information processing and quantum-limited measurements with superconducting circuits, where extremely low-noise microwave amplifiers are needed. 

We thank P.~Delsing for providing the device. We also thank C.~M.~Wilson and N.~Lambert for fruitful discussions. I.-C. H. acknowledge financial support from the MOST under project 104-2112-M-007-022-MY3 and MOST 104-2112-M-007-023, Taiwan. A.F.K. acknowledges support from a JSPS Postdoctoral Fellowship for Overseas Researchers P15750. F.N. acknowledges support from the RIKEN iTHES Project, the MURI Center for Dynamic Magneto-Optics via the AFOSR award number FA9550-14-1-0040, the Japan Society for the Promotion of Science (KAKENHI), the IMPACT program of JST, JSPS-RFBR grant No 17-52-50023, CREST grant No. JPMJCR1676, and the John Templeton Foundation. H. I. acknowledges support from FDCT of Macau under grants 013/2013/A1 and 065/2016/A2, University of Macau under grant MYRG2014-00052-FST, and National Natural Science Foundation of China under grant No. 11404415.
%

\end{document}